\documentclass[twocolumn,aps,prl,showpacs,amsmath,superscriptaddress,longbibliography,notitlepage]{revtex4-2}
\usepackage{times}
\usepackage{amssymb}
\usepackage{mathrsfs}
\usepackage{graphicx}
\usepackage{float}
\usepackage[caption=false]{subfig}
\usepackage[pdftex,colorlinks,linkcolor={red!75!black},citecolor={red!75!black},urlcolor={blue!75!black}]{hyperref}
\usepackage{verbatim}
\usepackage{epstopdf}
\usepackage[normalem]{ulem}
\usepackage{bm}
\usepackage{xcolor}
\usepackage{cancel}
\usepackage{soul}
\usepackage{tikz}
\usepackage{cancel}
\usepackage[normalem]{ulem} 
\usetikzlibrary{shapes.misc}
\usetikzlibrary{decorations.markings}

\newcommand{\clr}{\color{red!75!black}}

\tikzset{
	midin/.style={postaction={decorate},
		decoration={markings, mark=at position 0.55 with {\arrow{<}}}}, 
	midout/.style={postaction={decorate},
		decoration={markings, mark=at position 0.55 with {\arrow{>}}}}  
}

\begin{document}
	\title{Exceptional points and spectral cusps from density-wave fluctuation}
	\author{Zixi Fang}
	\affiliation{Beijing National Laboratory for Condensed Matter Physics, and Institute of Physics, Chinese Academy of Sciences, Beijing 100190, China}
	\affiliation{University of Chinese Academy of Sciences, Beijing 100049, China}
	
	\author{Chen Fang}
	\email{cfang@iphy.ac.cn}
	\affiliation{Beijing National Laboratory for Condensed Matter Physics, and Institute of Physics, Chinese Academy of Sciences, Beijing 100190, China}
	\affiliation{Kavli Institute for Theoretical Sciences, Chinese Academy of Sciences, Beijing 100190, China}
	
	\begin{abstract}
		We report two types of singularities that arise from fluctuations during the formation of charge- or spin-density waves.
		The first is the exceptional point (EP), corresponding to a higher-order pole of the retarded Green’s function.
		Such EPs lead to algebraic corrections in the decay of quasiparticle occupations and are observable through time-resolved angle-resolved photoemission spectroscopy (Tr-ARPES).
		The second is a spectral cusp, defined by the coalescence of three extrema in the real-frequency spectral function $A(\mathbf{k}, \omega)$.
		This cusp enforces the formation of Fermi arcs and induces a “threading” structure in the nearby band structure, both of which are directly observable in ARPES.
	\end{abstract}
	
	\maketitle
	
	\emph{\clr Introduction.---}
	Exceptional points (EPs) are non-diagonalizable points of a Hamiltonian at which both eigenvalues and eigenvectors coalesce \cite{Ganainy2018,huang2024acousticNatureReviewPhysics,scheibner2020odd,DingKun2023PRL,SongF2019_PRL,Liouvillian_Skin_PRL,Daley04032014,Yao2018,Kunst2018_PRL,WangZhong2018,Kai2022NC,Murakami2019_PRL,ChingHua2019,LeeCH2019_PRL,LonghiPRR2019,Kai2020,Okuma2020_PRL,Slager2020PRL,Zhesen2020_aGBZ,Zhesen2020_SE,XuePeng2020,Ghatak2020,Thomale2020,LiLH2020_NC,Kawabata2020_Symplectic,Wanjura2020_NC,XueWT2021_PRB,LiLH2021_NC,ZhangDDS2022,Longhi2022PRL,YMHu2022,Kai2022NC,Kawabata2024PRL,FuLiang2018_PRL}.
	Experimental observation of EPs has been confined to artificial systems \cite{DuanLM2017PRL,Zhou2018,ZhenBo2015,YuliPRB2013,san2016majoranaEP}, such as photonic and acoustic platforms, where non-Hermitian Hamiltonians are well controlled and both eigenvalues and eigenvectors are accessible. In condensed-matter systems, non-Hermiticity is encoded in the self-energy–corrected Green’s function, $G^R(k,\omega) = [\omega - h_0(k) - \Sigma(k,\omega)]^{-1}$. Previous studies have shown that disorder can split Dirac points into pairs of EPs \cite{Sato2019_EP,ZyuzinPRB2018,NoriPRL2021,SatoshiPRB2019}, and that strong correlations in Kondo lattices can similarly give rise to EPs \cite{FuLiang2020_PRL,yoshida2018PRB,Yoshida2021symmetryprotectedEP,FuLEP2024PRB}, by introducing asymmetric lifetimes for $d$- and $f$-electrons tunable by thermal fluctuations. In these models, EPs result from the competition between the hybridization and the dissipation of two bands. For example, in the Kondo insulator model, EPs appear when the hybridization energy scale matches the difference of the decay rates of $d$- and $f$-electrons. Motivated by these studies, we revisit a concept in condensed matter, spontaneous symmetry breaking driven by Fermi surface instability, where both hybridization and dissipation between electron bands naturally exist. The order parameter hybridizes electrons from different pieces of the Fermi surface, and the thermal fluctuation of the order parameter serves as a source of dissipation near the critical temperature \cite{QiYangPRB2023fluctuations}. This perspective raises our central question: can order-parameter fluctuations themselves serve as a primary mechanism for generating EPs in systems with spontaneous symmetry breaking, thereby broadening the range of condensed-matter platforms in which EPs may be realized?
	
	In this work, we investigate two types of singularities arising in the Brillouin zone (BZ) during the formation of charge/spin density wave (CDW/SDW) order, driven by order-parameter fluctuations. Modeling these fluctuations as spatially correlated disorder \cite{QiYangPRB2023fluctuations}, we compute the fluctuation-renormalized Green’s function to characterize their impact.
	The first singularity is the EP, where the retarded Green’s function develops a higher-order pole. While non-interacting Hermitian systems yield only simple poles, $G^R(k, \omega) \sim 1/(\omega - E(k))$, a second- or higher-order pole, $G^R(k, \omega) \sim 1/(\omega - E(k))^{n\geq 2}$, signals the emergence of an EP. If the interacting Green’s function $G^R(k, \omega)$ admits analytic continuation into the complex plane as a meromorphic function, the pole order provides a diagnostic for identifying EPs in correlated systems. Such higher-order poles alter the time-domain behavior as $G^R(k, t) \sim t^{n-1} e^{-iE(k)t}$, introducing anomalous algebraic corrections to the decay of quasiparticle occupations. These signatures can be probed via pump-probe spectroscopies, offering a route to experimentally access EPs in interacting quantum materials.
	The second singularity, identified here for the first time, is a real-frequency structure in the spectral function $A(\mathbf{k},\omega)$ that we term the spectral cusp. It originates from the coalescence of three extrema in $A(\mathbf{k}, \omega)$. This cusp is topologically protected, with the extrema of $A(\mathbf{k},\omega)$ forming a threading-like configuration in its vicinity. It gives rise to a Fermi arc and both the threading structure and arc are observable in ARPES spectra, thereby establishing the spectral cusp as a new class of topological singularity in non-Hermitian systems.
	
	To substantiate these findings, we implement the framework in a lattice model of iron-based superconductors, where spatially correlated SDW fluctuations are shown to induce both the EPs and the spectral cusps.
	
\begin{figure}[t]
		\centering
		\includegraphics[width=\linewidth]{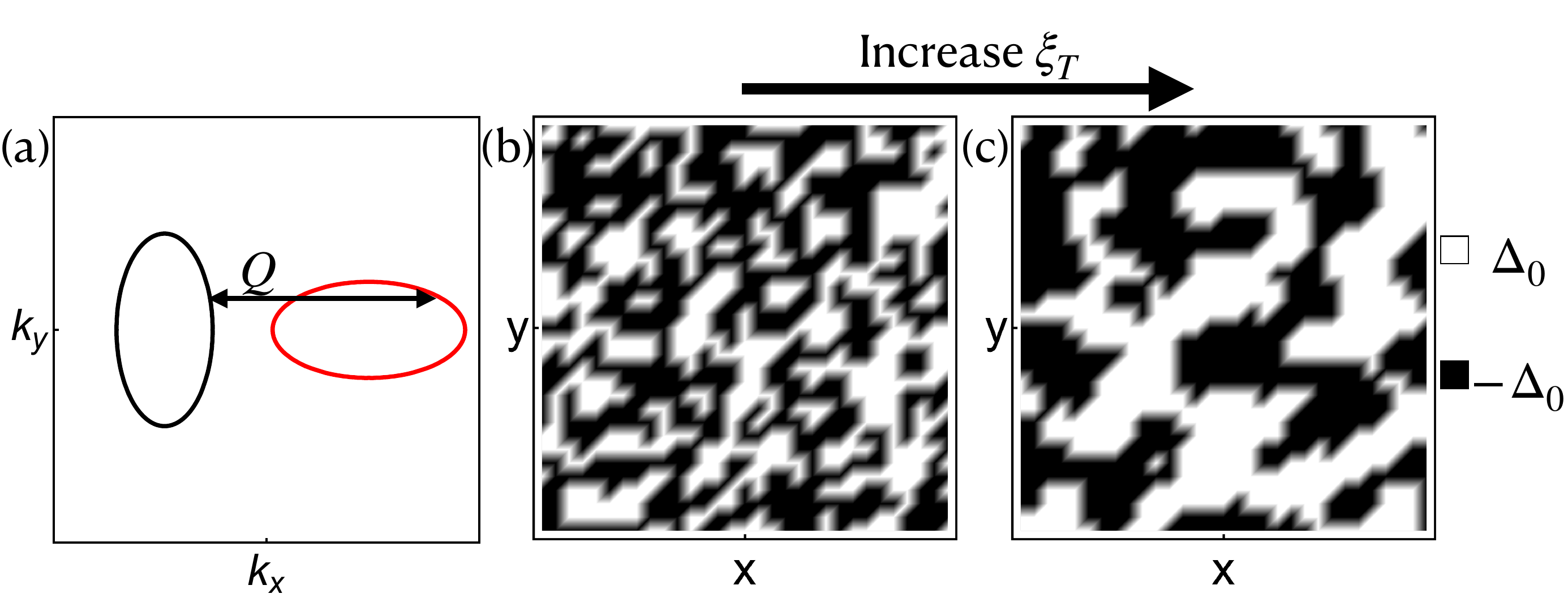}
		\caption{(a) Schematic Fermi surface consisting of one electron pocket (black) and one hole pocket (red), coupled by a density-wave order with ordering vector $\mathbf{Q}$. (b),(c) Real-space distributions of the density-wave order parameter $\Delta(\mathbf r)$. From (b) to (c), lowering temperature toward the transition enhances the correlation length $\xi_T$, leading to stronger spatial correlations.}
		\label{fig:1}
	\end{figure}

	\emph{\clr Model---} 
	We begin with a model Hamiltonian in which electrons are coupled to a fluctuating density-wave order, $\Delta(\mathbf{r})$. The Hamiltonian is written as
	\begin{align}
		\label{Eq:H}
		H\!=\!\int\!d\mathbf{k}(\epsilon_{\mathbf{k}}\!-\!\mu)c^\dagger_{\mathbf{k}} c_{\mathbf{k}}\!+\!\int d\mathbf{r} \left( e^{i\mathbf{Q}\cdot\mathbf{r}} \Delta(\mathbf r) c^\dagger_\mathbf{r}c_\mathbf{r} +h.c. \right).
	\end{align}
	The first term describes a single-band system with dispersion $\epsilon_{\mathbf{k}}$, featuring one electron-like Fermi pocket (black) and one hole-like Fermi pocket (red), as shown in Fig.~\ref{fig:1}(a).
	For simplicity, the electron and hole pockets are modeled by parabolic dispersions with opposite curvature,
	$\epsilon_{\mathbf{q}}^{(e,h)} = \pm(\frac{q_x^2}{m_{x,y}} + \frac{q_y^2}{m_{y,x}})$, near the respective pocket centers.
	The second term represents the coupling between electrons and the density-wave order parameter, with $\Delta(\mathbf{r})$ an Ising-like field encoding the amplitude of the density-wave order. In the ordered phase, $\Delta(\mathbf{r})$ is uniform; approaching the critical point from the disordered side, fluctuations render it spatially inhomogeneous with correlation length $\xi_T$. The correlation length $\xi_T$ serves as a measure of the spatial extent over which $\Delta(\mathbf{r})$ remains uniform.
	
	To simplify the discussion, we focus on spatial fluctuations of $\Delta(\mathbf{r})$ and neglect temporal ones
	\footnote{When temporal fluctuations are included, their effect is equivalent to a renormalization of the spatial correlation length, as discussed in detail in the Appendix.}. 
	$\Delta(\mathbf{r})$ is modeled as a static, spatially correlated disorder characterized by
	$\overline{\Delta(\mathbf{r})}=0,\quad
	\overline{\Delta(\mathbf{r}_i)\Delta(\mathbf{r}_j)}=\Delta_0^2 e^{-|\mathbf{r}_i-\mathbf{r}_j|^2/\xi_T^2}$,
	where $\Delta_0$ denotes the fluctuation strength, and the overline represents ensemble averaging.
	
	Intuitively, in the disordered phase close to $T_c$ ($T \to T_c^+$), thermal fluctuations generate correlated domains of size $\xi_T$, as illustrated in Fig.~\ref{fig:1}(b) and (c). Within each domain, $\Delta(\mathbf r)$ is uniform, but it varies across domains. Translational symmetry is thus locally broken but statistically restored upon ensemble averaging~\cite{QiYangPRB2023fluctuations}. This domain picture clarifies why the disorder description above captures the essential effects of order-parameter fluctuations.
	
	In this ensemble-averaged picture, we treat $\Delta_0$ as a perturbative parameter and retain only the leading-order self-energy correction to the retarded Green’s function $G^R(\mathbf{k}, \omega)= (\omega - \epsilon_\mathbf{k} - \Sigma(\mathbf{k}, \omega))^{-1}$ \cite{SupMat}:
	\begin{align}
		\label{eq:SelfEnergy}
		\Sigma(\mathbf{k}, \omega) &= \frac{\Delta_0^2}{\omega - \epsilon_{\mathbf{k} + \mathbf{Q}} + i \gamma_{\mathbf{k} + \mathbf{Q}}(T)},
	\end{align}
	Compared to the self-energy induced by static density-wave order, the fluctuation-induced self-energy acquires an additional imaginary part, $\gamma_\mathbf{k}(T) = v_\mathbf{k}/\xi_T$, with $v_\mathbf{k}=|\nabla \epsilon_\mathbf{k}|$ the group velocity at $\mathbf{k}$, thereby giving quasiparticles a finite lifetime. As $T \to T_c^+$, the correlation length $\xi_T$ diverges, rendering $\gamma_\mathbf k$ a non-Hermitian term tunable by temperature. Since the poles of $G^R$ encode the non-Hermitian spectrum, varying $\xi_T$ controls its structure, thereby making the emergence of EPs possible in this system. 
	
	\begin{figure}[t]
		\centering
		\includegraphics[width=\linewidth]{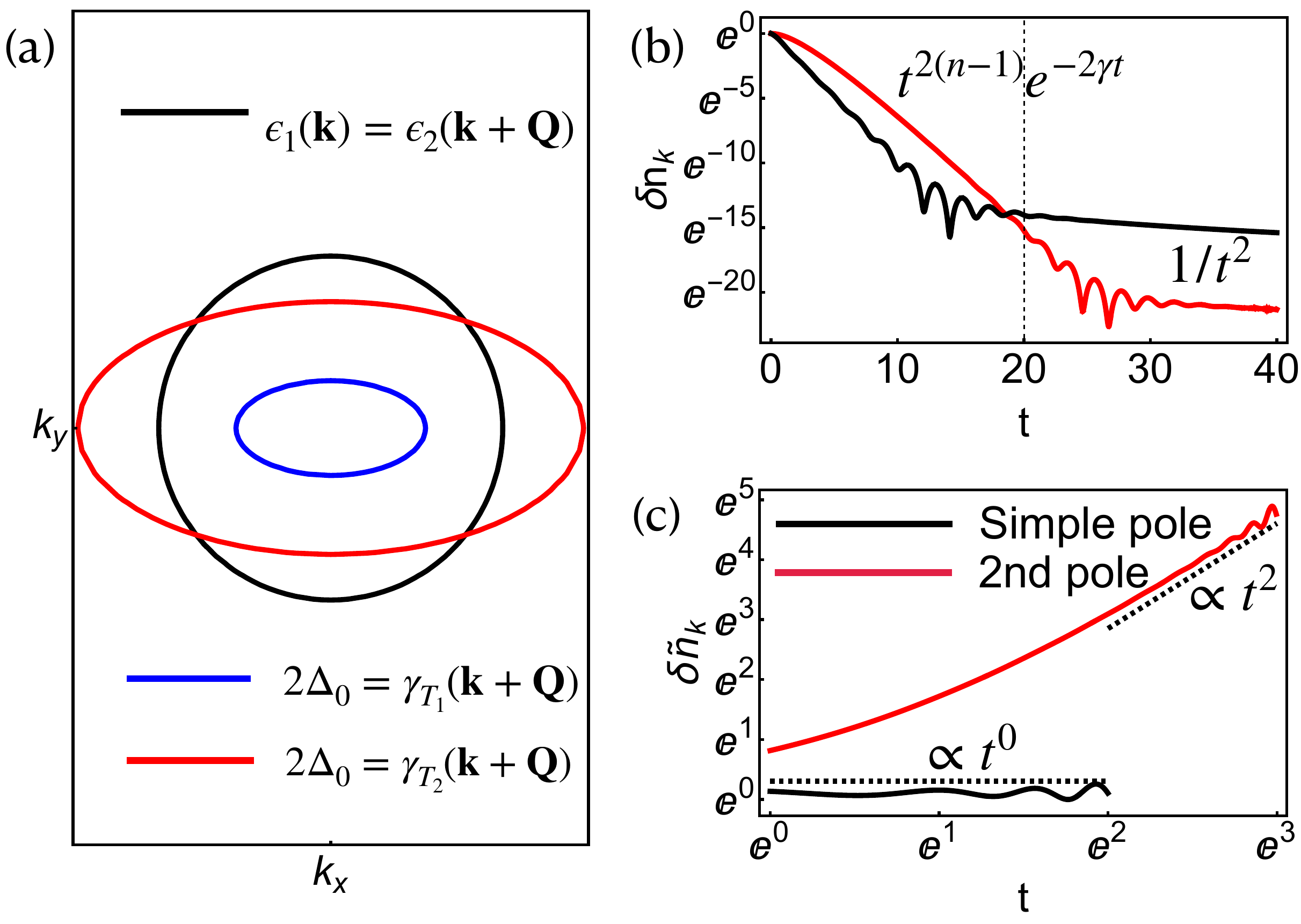}
		\caption{(a) Conditions for emergence of EPs. The black line corresponds to Eq.\ref{Eq:DpLine}, while the colored lines correspond to Eq.\ref{Eq:DeltaEqgamma}; the red curve represents a lower temperature (larger $\xi_T$) than the blue one.
			(b) and (c) Dynamical signatures of a simple pole and a second-order pole. Panel (c) shows the doubly logarithmic plot after applying the exponential correction $\delta \tilde{n}_{\mathbf k} = e^{2\mathrm{Im}\omega_{\mathbf k}t}\delta n_{\mathbf k}$, highlighting the algebraic behavior.}
		\label{fig:2}
	\end{figure}
	\emph{\clr Exceptional Points---}
	To analyze these EPs explicitly, we now examine the poles of the retarded Green’s function $G^R(\mathbf{k},\omega)$, which define the quasiparticle spectrum. Inserting Eq.~\ref{eq:SelfEnergy} into $G^R(\mathbf{k},\omega)$, these poles take the form
	$\omega_\pm(\mathbf{k}) =  \frac{\epsilon_\mathbf{k} + \epsilon_{\mathbf{k+Q}} - 2i\gamma_{\mathbf{k} + \mathbf{Q}}(T)}{2} \pm \sqrt{ \frac{ \left( \epsilon_\mathbf{k} - \epsilon_{\mathbf{k+Q}} + 2i\gamma_{\mathbf{k} + \mathbf{Q}}(T) \right)^2 }{4} + \Delta_0^2 }.$
	These two poles suggest an apparent band-splitting at each momentum $\mathbf{k}$, mimicking a two-band structure.
	
	When two poles meet at $\mathbf k^*$, $\omega_{+,\mathbf k^*}=\omega_{-,\mathbf k^*}$, we define the winding
	$\nu(\mathbf k^*)=-\frac{1}{2\pi}\oint_{\Gamma(\mathbf k^*)}\nabla_{\mathbf k}\arg[\omega_{+,\mathbf k}-\omega_{-,\mathbf k}]\cdot d\mathbf k$,
	with $\Gamma(\mathbf k^*)$ a closed loop around $\mathbf k^*$. A nonzero $\nu(\mathbf k^*)$ signals an EP. Since these poles are accessible through $G^R(\mathbf k,\omega)$, this prompts the question of how to distinguish an EP from an ordinary degeneracy.
	
	The key lies in the analytic structure of $G^R$. 
	In non-interacting Hermitian systems, the Hamiltonian is diagonalizable, so $G^R$ has only simple poles, $G^R(\mathbf k,\omega)\sim 1/(\omega-E_{\mathbf k})$. 
	By contrast, higher-order poles arise only when the Hamiltonian becomes non-diagonalizable—precisely at an EP~\cite{Heiss2012}. 
	Interactions, through the self-energy, can promote the order of poles. 
	Hence, the order of poles of $G^R$ identifies EPs in the spectrum.
	
	A higher-order pole in the retarded Green’s function arises when the following conditions are satisfied:
	\begin{align}
		G^R(\mathbf{k}, \omega)^{-1} = 0, \quad \partial_\omega G^R(\mathbf{k}, \omega)^{-1} = 0.
	\end{align}
	These impose four real constraints on the system’s parameters. Meanwhile, the number of tunable real degrees of freedom is $d+2$, with $d$ from momentum and $2$ from the complex frequency $\omega$. Therefore, second-order EPs can stably exist only in systems with spatial dimension $d\geq 2$, consistent with the conclusion in Ref.~\cite{YangFermiDoubling2021}.
	
	\begin{figure*}[htbp]
		\centering
		\includegraphics[width=\linewidth]{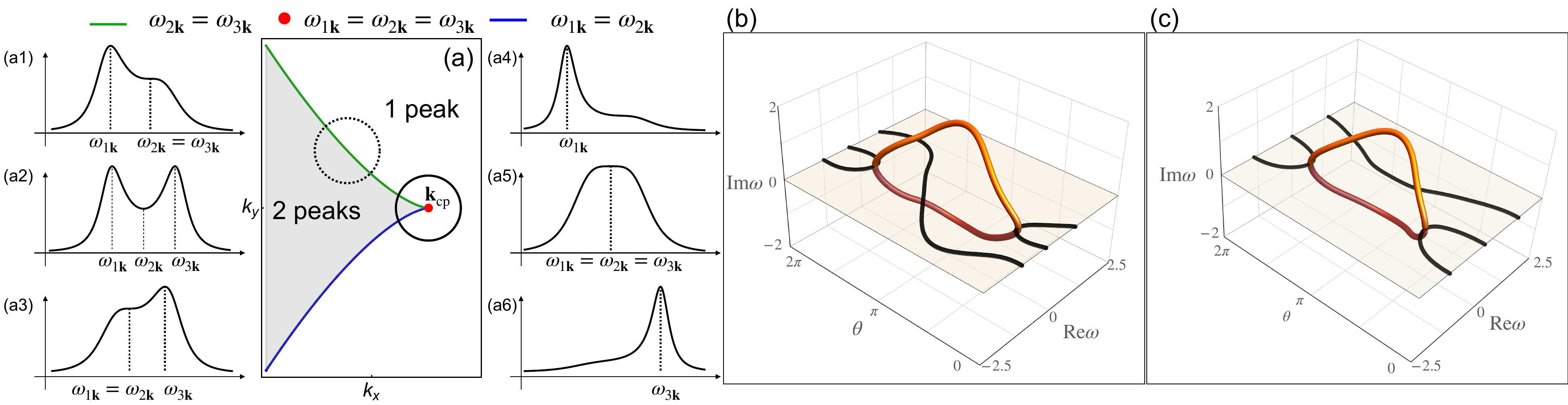}
		\caption{(a) Origin of the cusp. Green and blue lines indicate degeneracies $\omega_1=\omega_2$ and $\omega_2=\omega_3$; their intersection (red dot) marks the cusp. In the gray region, $A(\mathbf{k},\omega)$ exhibits two peaks (three extrema), while in the white region only a single peak remains.
			(a1)–(a6) Representative $A(\mathbf{k},\omega)$ at selected $\mathbf{k}$ points in (a). Panels (a1), (a3), and (a5) correspond to momenta on the green line, blue line, and at the cusp. Panel (a2) shows the two-peak regime; (a4) and (a6) the single-peak regime.
			(b),(c) Trajectories of spectral extrema $\omega_{i\mathbf{k}}$ along loops that either enclose (solid) or avoid (dashed) the cusp, as marked in (a). The colored surface denotes the real-frequency plane. Black curves lie within it and correspond to observable extrema in $A(\mathbf{k},\omega)$, while orange curves depart from it, indicating annihilated extrema beyond the real-frequency plane.}
		\label{fig:3}
	\end{figure*}
	
	Substituting Eq.~\ref{eq:SelfEnergy} into this criterion, the condition for realizing a second-order EP becomes:
	\begin{align}
		\label{Eq:DpLine}
		\epsilon_\mathbf{k} &= \epsilon_{\mathbf{k} + \mathbf{Q}}, \\
		\label{Eq:DeltaEqgamma}
		2\Delta_0 \xi_T &= v_{\mathbf{k} + \mathbf{Q}}(T).
	\end{align}
	These two equations define distinct curves in momentum space: Eq.~\ref{Eq:DpLine} identifies points of real-part band degeneracy, while Eq.~\ref{Eq:DeltaEqgamma} imposes a constraint on the strength of fluctuations. As shown in Fig.~\ref{fig:2}(a), the gray dashed line corresponds to Eq.~\ref{Eq:DpLine} and the pink dashed line to Eq.~\ref{Eq:DeltaEqgamma}; their intersections (red dots) mark the locations of EPs in the BZ.
	
	When $T \to T_c^+$, the correlation length diverges as $\xi_T \propto |T-T_c|^{-\nu}$ and $\Delta_0$ remains finite, where $\nu=0.5$ is the critical exponent in mean-field theory.
	Consequently, the extent of the Eq.~\ref{Eq:DeltaEqgamma} curve is governed by the correlation length $\Delta_0\xi_T\propto \xi_T$: it starts from a single point near the band top/bottom at $T\rightarrow\infty$ and sweeps the entire BZ as $T$ approaches $T_c$. In contrast, Eq.~\ref{Eq:DpLine} is independent of $\xi_T$. As a result, over a finite range of temperature, the two curves intersect, ensuring the presence of EPs in the system.
	
	\emph{\clr The anomalous decay at exceptional points---}
	To probe the effect of a higher-order pole, we focus on the quasiparticle dynamics. We consider a weak ultrafast pump pulse that excites the system from its ground state $|\psi_0\rangle$ to a superposition $|\Psi\rangle = \sqrt{1-a^2}|\psi_0\rangle + a|\psi_{\text{eh}}\rangle$, with $a \ll 1$. Here $|\psi_{\text{eh}}\rangle = \sum_{\mathbf{k}} Y_{\mathbf{k}} c^\dagger_{\mathbf{k} + \mathbf{q}_p} c_{\mathbf{k}} |\psi_0\rangle$ describes a single electron-hole excitation with momentum transfer $\mathbf{q}_p$, and $Y_{\mathbf{k}}$ is the effective excitation strength arising from the light–matter coupling. The subsequent time evolution is governed by $|\Psi(t)\rangle = e^{-iHt} |\Psi\rangle$.
	
	We analyze the momentum-resolved occupation $n_\mathbf{k}(t) = \langle \Psi(t)|c_\mathbf{k}^\dagger c_\mathbf{k}|\Psi(t)\rangle$.
	To leading order in the pump amplitude, its time-dependent part reads
	\begin{align}
		\label{Eq:decay}
		\begin{aligned}
			\delta n_\mathbf{k}(t)=& |Y_{\mathbf{k}-\mathbf q_p} a|^2  n_0(\mathbf{k} - \mathbf{q}_p) |G^>(\mathbf{k}, t)|^2 \\
			& - |Y_{\mathbf{k}} a|^2(1 - n_0(\mathbf{k} + \mathbf{q}_p)) |G^<(\mathbf{k}, t)|^2,
		\end{aligned}
	\end{align}
	where $n_0(\mathbf{k}) = \langle \psi_0 | c_\mathbf{k}^\dagger c_\mathbf{k} | \psi_0 \rangle$ is the ground-state occupation. The Green’s functions $G^<(\mathbf{k}, t)=\langle \psi_0| e^{iHt}c_\mathbf{k}^\dagger e^{-iHt} c_\mathbf{k} |\psi_0\rangle$ and $G^>(\mathbf{k}, t)=\langle \psi_0| c_\mathbf{k} e^{iHt}c_\mathbf{k}^\dagger e^{-iHt}  |\psi_0\rangle$ encode the post-pump relaxation dynamics and carry signatures of EPs.
	The first term in Eq.~\ref{Eq:decay} corresponds to an electron excitation: a photon with momentum $\mathbf{q}_p$ excites an electron from a filled state at $\mathbf{k} - \mathbf{q}_p$ to an unoccupied state at $\mathbf{k}$. The second term describes a hole excitation: an electron at $\mathbf{k}$ is promoted to $\mathbf{k} + \mathbf{q}_p$, leaving behind a hole at $\mathbf{k}$.
	
	For concreteness, we focus on the case where the EP lies above the Fermi surface, such that the hole contribution is negligible. We then retain only the first term, yielding $\delta n_\mathbf{k}(t) \propto |G^>(\mathbf{k}, t)|^2$, which probes the time-domain signature of the quasiparticle excitation.
	To isolate the dynamical imprint of EPs, we examine the long-time behavior of the excited-state occupation,
	\begin{align}
		\label{eq: quasiparticle_decay}
		\delta n_\mathbf{k}(t)\! =\! \left|\! \int\!\! d\omega A(\mathbf{k}, \omega) e^{-i\omega t}\! +\! \frac{A(\mathbf{k}, \mu) e^{-i\mu t}}{it}\! +\! O\!\!\left(\frac{1}{t^2}\right)\! \right|^2\!\!\!\!,
	\end{align}
	where $A(\mathbf{k}, \omega) = -\mathrm{Im} \left[ G^R(\mathbf k,\omega) \right]$ is the spectral function associated with an n-th order pole at $\omega_\mathbf{k}$, and the second term represents a subleading correction from the Fermi surface. When the pole is far from the Fermi energy, we have $A(\mathbf{k},\mu)\rightarrow0$, and the quasiparticle decay follows $\delta n_\mathbf{k}(t) \sim t^{2(n-1)} e^{2\mathrm{Im}\,\omega_\mathbf{k} t}(t\gg 1)$, where the power-law prefactor $t^{2(n-1)}$ reflects the order-$n$ nature of the Green’s function pole. This behavior marks a distinction: while a simple pole ($n=1$) yields pure exponential decay, higher-order poles produce an additional algebraic component. 
	In the long-time limit, the second term in Eq.~(\ref{eq: quasiparticle_decay}) inevitably dominates $\delta{n}_{\mathbf{k}}(t)$ and shows an algebraic tail $1/t^2$. In Fig.\ref{fig:2}(c), we identify the time window in which the exponential decay is dominant, and fit the simulated $\delta{n}_\mathbf{k}(t)$ for the order of the pole.
	
	\emph{\clr Spectral cusps---}
	Beyond the quasiparticle dynamics, condensed-matter experiments also measure the spectral function $A(\mathbf{k},\omega\in\mathbb{R})$, from which information about the poles is typically extracted by assuming a specific form for $A(\mathbf{k},\omega)$. The conventional approach assumes simple poles in $G^R$, giving Lorentzian $A(\mathbf{k},\omega)$. 
	In contrast, our analysis shows that $G^R$ can host richer structures, including higher-order poles at EPs, where $A(\mathbf{k},\omega)$ becomes non-Lorentzian. Without any prior assumption about $G^R$, what robust information can be obtained directly from $A(\mathbf{k},\omega\in\mathbb{R})$?
	
	In a two-band system, $A(\mathbf{k}, \omega)$ typically displays two broadened peaks separated by a dip, as illustrated in Fig.\ref{fig:3}(a2). However, peak broadening can obscure the two-peak structure: even when the poles of $G^R$ remain distinct, only a single peak may be visible in $A(\mathbf{k},\omega)$, as illustrated by the evolution from Fig.~\ref{fig:3}(a2) to (a1) and then to (a4), where unequal quasiparticle lifetimes cause one peak $\omega_{3\mathbf{k}}$ and one dip $\omega_{2\mathbf{k}}$ to coincide and annihilate. 
	Consequently, an extended region appears in momentum space where $A(\mathbf{k},\omega)$ displays only one peak.
	The boundary of this region, denoted by $\mathcal{B}$, corresponds to the condition where a peak and a dip merge, $\partial_\omega A(\mathbf{k},\omega)=\partial_\omega^2 A(\mathbf{k},\omega)=0$, marking the transition between the one-peak and two-peak regimes, as shown by colored line in Fig.~\ref{fig:3}(a).
	
	At generic points, the boundary curve $\mathcal{B}$ is smooth. However, at certain isolated momenta $\mathbf{k}_{\mathrm{cp}}$, $\mathcal{B}$ develops a cusp, marked by the red dot in Fig.~\ref{fig:3}(a).
	At such a cusp, all three extrema of $A(\mathbf{k},\omega)$ coalesce, defined by
	\begin{align}
		\label{eq:CuspCondition}
		\partial_\omega A = \partial_\omega^2 A = \partial_\omega^3 A = 0.
	\end{align}
	Since these impose three constraints in the three-dimensional $(k_x, k_y, \omega)$ space, $\mathbf{k}_{\mathrm{cp}}$ is topologically protected and persists across a finite temperature range.
	Near $\mathbf{k}_{\mathrm{cp}}$, $A(\mathbf{k},\omega)$ contains three extrema (two peaks and one dip), allowing $\partial_\omega A$ to be expanded cubically around $\omega = E_{\mathrm{cp}}$ as $\partial_\omega A = \frac{\partial_\omega^4 A(\mathbf{k},E_{\mathrm{cp}})}{6} \left(\delta\omega^3 + a_{\mathbf{k},T}\delta\omega + b_{\mathbf{k},T}\right)$, where $\delta\omega = \omega - E_{\mathrm{cp}}-\partial_\omega^3 A(\mathbf{k}, E_{\mathrm{cp}})/\partial_\omega^4 A(\mathbf{k}, E_{\mathrm{cp}})$, and $a_{\mathbf{k},T}, b_{\mathbf{k},T}$ vary smoothly with $\mathbf{k}$ and $T$
	\footnote{Since the equation $\partial_\omega A(\mathbf{k}, \omega)=0$ admits three roots near $\mathbf{k}=\mathbf{k}_{\mathrm{cp}},\omega=E_{\mathrm{cp}}$, it must locally take the form of a cubic equation in $\omega$. Expanding $\partial_\omega A$ around $\omega = E_{\mathrm{cp}}$ yields
		$\partial_\omega A = \alpha_{\mathbf k}^{(1)} + \alpha_{\mathbf k}^{(2)}(\omega - E_{\mathrm{cp}})
		+\frac{\alpha_{\mathbf k}^{(3)}}{2}(\omega - E_{\mathrm{cp}})^2
		+\frac{\alpha_{\mathbf k}^{(4)}}{6}(\omega - E_{\mathrm{cp}})^3$,
		where $\alpha_{\mathbf k}^{(n)} = \partial_\omega^n A(\mathbf{k}, E_{\mathrm{cp}})$.
		By shifting the frequency variable as
		$\delta\omega = \omega - E_{\mathrm{cp}} - \frac{\alpha_{\mathbf k}^{(3)}}{\alpha_{\mathbf k}^{(4)}}$,
		this cubic can be reduced to the canonical form used in the main text, with $a_{\mathbf{k},T}=6\alpha_{\mathbf k}^{(2)}-3(\alpha_{\mathbf k}^{(3)})^2$ and $b_{\mathbf{k},T}=2(\alpha_{\mathbf k}^{(3)})^3+6\alpha_{\mathbf k}^{(3)}\alpha_{\mathbf k}^{(2)}+6\alpha_{\mathbf k}^{(1)}$.}.
	The boundary $\mathcal{B}$ corresponds to condition that a peak and a dip merge, given by the vanishing discriminant
	$\mathrm{Disc}[\partial_\omega A] = -\left(4a_{\mathbf{k},T}^3 + 27b_{\mathbf{k},T}^2\right) = 0$. 
	This defines an elliptic curve that has exactly one cusp singularity at $a_{\mathbf{k},T}=b_{\mathbf{k},T}=0$. When $a_{\mathbf{k},T} \approx \mathbf{v}_a \cdot(\mathbf{k}-\mathbf{k}_{\mathrm{cp}})$ and $b_{\mathbf{k},T} \approx \mathbf{v}_b \cdot (\mathbf{k}-\mathbf{k}_{\mathrm{cp}})$ vary linearly with momentum, this cusp maps directly onto the cusp in momentum space, marked by red dot in Fig.~\ref{fig:3}(a).
	
	Along the boundary curve $\mathcal{B}$, $\mathbf{k}_{\mathrm{cp}}$ is a branching point, and the two branches correspond to distinct types of extrema coalescence: for one branch above $\mathbf{k}_{\mathrm{cp}}$, $\omega_2 = \omega_3$ [green line in Fig.~\ref{fig:3}(a)], while for the other branch below $\mathbf{k}_{\mathrm{cp}}$, $\omega_1 = \omega_2$ (blue line). Encircling $\mathbf{k}_{\mathrm{cp}}$ in momentum space produces a threading-like evolution of extrema. As illustrated in Fig.\ref{fig:3}(b), a peak and a dip meet and annihilate, causing their corresponding extrema to leave the real-frequency axis before reappearing (orange trajectory). 
	The third peak $\omega_3$ passes through the orange loop and returns to where $\omega_1$ was. This threading process effectively exchanges $\omega_{1,3}$. 
	In contrast, when the path avoids the cusp (Fig.\ref{fig:3}(c)), no such threading occurs. This confirms that the cusp represents a topologically protected singularity of the spectral function.
	
	Furthermore, we find that the presence of a cusp necessarily implies the emergence of a Fermi arc in the band structure defined by the spectral peak of $A(\mathbf{k},\omega)$. The cusp serves precisely as the endpoint of this arc. A detailed proof is provided in the Appendix. In summary, the spectral cusp identified here represents a new type of topological singularity in the momentum-resolved spectral function, which also provides a mathematical origin for the Fermi arc observable in $A(\mathbf{k},\omega)$.
	
	\begin{figure}[t]
		\centering
		\includegraphics[width=\linewidth]{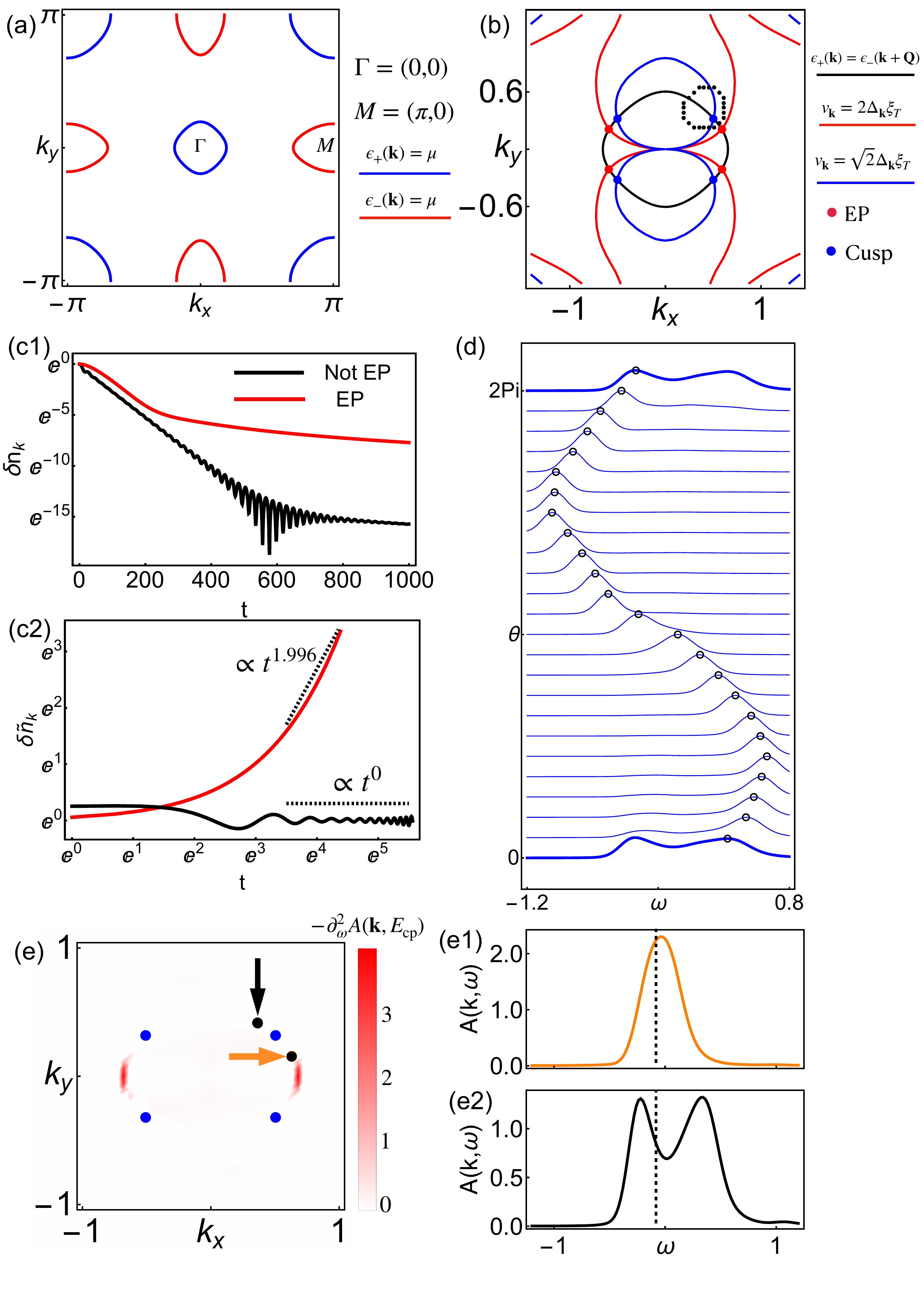}
		\caption{Simulation of the model in Eq.~\ref{Eq:LatticeModel} with parameters $(t_1, t_2, t_3, t_4, \mu, \xi\Delta_0)=(-0.9, 1.4, -0.85, -0.85, 1.45, 0.92)$.
			(a) Fermi surface: blue lines denote electron-like pockets and red lines denote hole-like pockets, with nesting occurring between those around $\Gamma=(0,0)$ and $M=(\pi,0)$.
			(b) Conditions for EP (red dots) and cusp (blue dots) formation.
			(c) Dynamical response with $\Delta_0=0.1$ at EP and a generic momentum ($\mathbf{k}=(0.2,0.5)$) away from EP. Panel (c2) shows the exponential correction $\delta \tilde{n}_{\mathbf k} = e^{2\mathrm{Im}\omega_{\mathbf k}t}\delta n_{\mathbf k}$, highlighting the algebraic behavior.
			(d) and (e) Simulation of the spectral function on a $120 \times 120$ lattice with $\Delta_0=0.5$.
			(d) Evolution of $A(\mathbf{k},\omega)$ along a closed loop around the cusp (black dotted path in panel (b)).
			(e) Intensity map of $-\partial_\omega^2A(\mathbf k,E_\mathrm{cp})$.
			(e1), (e2) Spectral functions at points indicated by the orange and red arrows in (e); dashed lines mark $\omega = E_\mathrm{cp}$.
		}
		\label{fig:4}
	\end{figure}
	
	\emph{\clr Lattice Simulation---} 
	We test our theory in a two-band lattice model for iron-based superconductors~\cite{Shou-ChengSDWPRB2008}. The tight-binding Hamiltonian is
	\begin{align}
		\label{Eq:LatticeModel}
		H_0=(d_{x\mathbf{k}}^\dagger, d_{y\mathbf{k}}^\dagger)
		\left(
		\begin{matrix}
			\varepsilon_{x\mathbf{k}} -\mu& \varepsilon_{x y\mathbf{k}}\\
			\varepsilon_{x y\mathbf{k}}&\varepsilon_{y\mathbf{k}}-\mu
		\end{matrix}
		\right)
		\left(
		\begin{aligned}
			d_{x\mathbf{k}}\\
			d_{y\mathbf{k}}
		\end{aligned}
		\right).
	\end{align}
	Here $d_{x\mathbf{k}}$ and $d_{y\mathbf{k}}$ are annihilation operators for electrons in the $d_{xz}$ and $d_{yz}$ orbitals, respectively, with dispersions $\varepsilon_{x \mathbf{k}}=-2 t_{1} \cos k_x-2 t_{2} \cos k_y-4 t_3 \cos k_x \cos k_y$ and $\varepsilon_{y \mathbf{k}}=-2 t_{2} \cos k_x-2 t_{1} \cos k_y-4 t_3 \cos k_x \cos k_y$. 
	The orbital hybridization $\varepsilon_{xy\mathbf{k}}=-4t_4\sin k_x \sin k_y$ opens a gap between the orbitals, yielding the hybridized bands $\epsilon_{ \pm,\mathbf{k}}=\frac{\varepsilon_{x\mathbf{k}}+\varepsilon_{y\mathbf{k}}}{2} \pm \sqrt{(\frac{\varepsilon_{x\mathbf{k}}-\varepsilon_{y\mathbf{k}}}{2})^2+\varepsilon_{x y\mathbf{k}}^2}-\mu$, with Fermi surfaces shown in Fig.~\ref{fig:4}(a). 
	Interactions favor a commensurate SDW with $\mathbf Q=(\pi,0)$~\cite{Shou-ChengSDWPRB2008}. Fluctuations are modeled as correlated disorder, $V = \int d\mathbf{r} \left( e^{i\mathbf{Q}\cdot\mathbf{r}} \Delta(\mathbf{r})\, (d^\dagger_{x\mathbf{r}} d_{x\mathbf{r}} + d^\dagger_{y\mathbf{r}} d_{y\mathbf{r}})+h.c. \right)$. Multi-orbital mixing makes the coupling between $\epsilon_{-,\mathbf k}$ and $\epsilon_{+,\mathbf{k+Q}}$ momentum dependent: $\Delta_\mathbf{k}=\Delta_0 \sin(\frac{\theta_\mathbf{k}-\theta_{\mathbf{k+Q}}}{2})$, with $\sin\theta_{\mathbf{k}}=\frac{\varepsilon_{xy\mathbf{k}}}{\sqrt{\varepsilon_{xy\mathbf{k}}^2+(\varepsilon_{x \mathbf{k}}-\varepsilon_{y \mathbf{k}})^2/4}}$.
	
	Fig.~\ref{fig:4}(b) summarizes the criteria for EPs and spectral cusps. The cusp condition reduces to the intersection of $\sqrt{2}\Delta_{\mathbf k}=\gamma_{\mathbf{k}+\mathbf Q}(T)$ with Eq.~\ref{Eq:DpLine}. Fig.~\ref{fig:4}(c1,c2) compare the dynamics at the EP and at a nearby generic momentum: we first fit the exponential decay in Fig.~\ref{fig:4}(c1); subtracting it exposes in Fig.~\ref{fig:4}(c2) an algebraic tail unique to the EP. Along a closed loop encircling the cusp, tracking a single spectral peak reveals a branch exchange even though the initial and final spectra coincide, uncovering the cusp’s topology and providing an observable threading signature [Fig.~\ref{fig:4}(d)]. At $E_{\mathrm{cp}}$, plotting $-\partial_\omega^2 A(\mathbf k,E_{\mathrm{cp}})$ highlights extrema so that peaks appear as positive features; the Fermi surface then emerges as a red arc terminating at the blue cusp [Fig.~\ref{fig:4}(e)]. Two representative momenta confirm this: on the arc (orange arrow) the spectrum peaks at $E_\mathrm{cp}$ [Fig.~\ref{fig:4}(e1)], whereas off the arc (black arrow) it dips at $E_\mathrm{cp}$ [Fig.~\ref{fig:4}(e2)], demonstrating that the cusp truncates the Fermi surface and generates the arc.
	
	Our lattice implementation demonstrates that fluctuation-induced non-Hermiticity gives rise to two singularities: EPs with algebraic dynamics observed in tr-ARPES, and spectral cusps manifested in ARPES as cusp-terminated Fermi arcs with branch exchange.
	
	\bibliography{Refs_MainText}

	\appendix
	\setcounter{equation}{0}  
	\setcounter{figure}{0}  
	\renewcommand{\thefigure}{A\arabic{figure}}
	\renewcommand{\theequation}{A\arabic{equation}}
	\begin{widetext}
		\section{Appendix A. The cusp induced Fermi Arc}
		\begin{figure}[h]
			\centering
			\includegraphics[width=\linewidth]{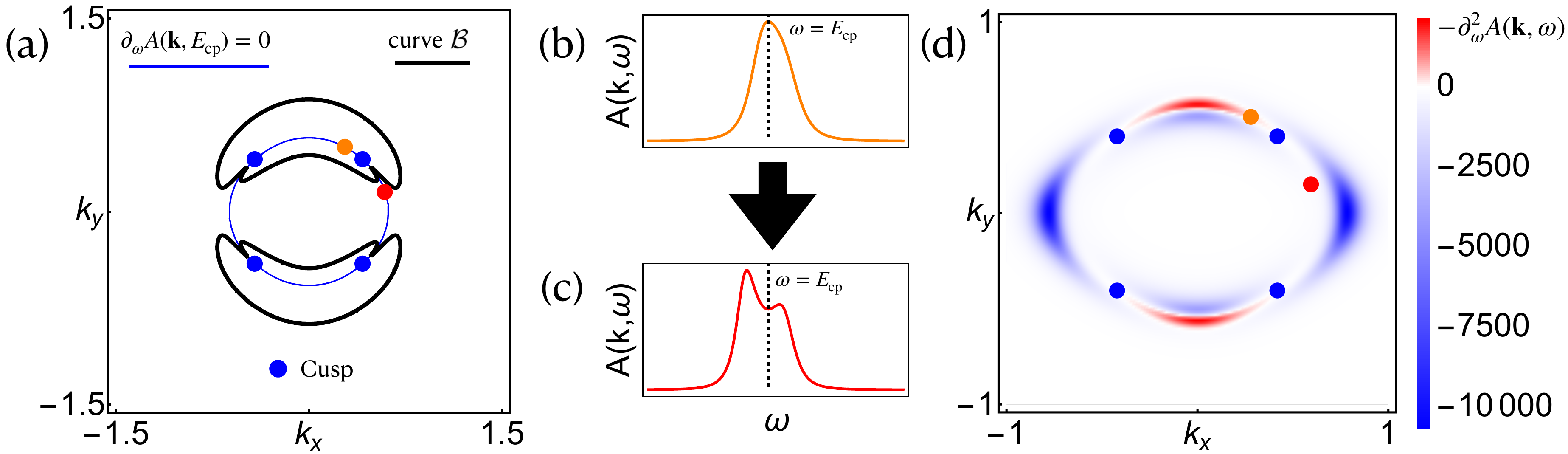}
			\caption{Illustration based on the model in Eq.~\ref{Eq:LatticeModel} with parameters $(t_1, t_2, t_3, t_4, \mu, \Delta_0, \xi)=(-0.9,,1.4,,-0.85,,-0.85,,1.45,,0.5,,1.845)$.
				(a) Schematic of the cusp: the blue curve denotes the possible Fermi surface (with Fermi arcs forming its subsets), and the black line indicates the condition where a dip and a peak coincide.
				(b),(c) Spectral functions at the black and red points in panels (a) and (d).
				The dashed line indicates $\omega = E_\mathrm{cp}$.
				(d) Intensity plot of the second derivative of the spectral function with respect to frequency.}
			\label{figsm:1}
		\end{figure}
		
		Near the cusp, the extremum condition for the spectral function, $\partial_\omega A(\mathbf{k},\omega)=0$, can be expanded to cubic order around $E_{\mathrm{cp}}$:
		$$\partial_\omega A = \alpha_{\mathbf k,1} + \alpha_{\mathbf k,2}(\omega - E_{\mathrm{cp}})
		+\frac{\alpha_{\mathbf k,3}}{2}(\omega - E_{\mathrm{cp}})^2
		+\frac{\alpha_{\mathbf k,4}}{6}(\omega - E_{\mathrm{cp}})^3$$
		where $\alpha_{\mathbf{k},n} = \partial_\omega^n A(\mathbf{k},E_{\mathrm{cp}})$.
		Fixing the Fermi level at $E_{\mathrm{cp}}$ imposes $\alpha_{\mathbf{k},1}=0$ [blue line in Fig.~S1(a)], and the extremum condition reduces to
		$$(\omega-E_{\mathrm{cp}}) \left(\alpha_{\mathbf k,2}
		+\frac{\alpha_{\mathbf k,3}}{2}(\omega - E_{\mathrm{cp}})
		+\frac{\alpha_{\mathbf k,4}}{6}(\omega - E_{\mathrm{cp}})^2\right)=0,$$
		which always yields one extremum pinned at $\omega = E_{\mathrm{cp}}$.
		The discriminant of this cubic equation is then $\mathrm{Disc}[\partial_\omega A]=\frac{\alpha_{\mathbf k,3}^2}{4}-\frac{2\alpha_{\mathbf k,2}\alpha_{\mathbf k,4}}{3}$
		At the cusp $\mathbf{k}_{\mathrm{cp}}$, we have $\alpha_{\mathbf{k}_{\mathrm{cp}},1}=\alpha_{\mathbf{k}_{\mathrm{cp}},2}=\alpha_{\mathbf{k}_{\mathrm{cp}},3}=0$.
		Expanding near $\mathbf{k}_{\mathrm{cp}}$ gives $\alpha_{\mathbf{k}_\mathrm{cp}+\mathbf{q},2} \approx \mathbf{v}_2 \cdot \mathbf{q}$ and $\alpha_{\mathbf{k}_{\mathrm{cp}}+\mathbf{q},3} \approx \mathbf{v}_3\cdot \mathbf{q}$, leading to $\mathrm{Disc}[\partial_\omega A]=-\frac{2\alpha_{\mathbf{k}_\mathrm{cp},4}\mathbf{v}_2\cdot \mathbf{q}}{3}\approx-\frac{2\alpha_{\mathbf k,2}\alpha_{\mathbf k,4}}{3}$.		
		On the side where $\mathrm{Disc}[\partial_\omega A]<0$, only a single peak remains [Fig.~S1(b)]. Crossing into the $\mathrm{Disc}[\partial_\omega A]>0$ region yields three extrema: one fixed at $E_{\mathrm{cp}}$, and two additional solutions to $\alpha_{\mathbf k,2}
		+\frac{\alpha_{\mathbf k,3}}{2}(\omega - E_{\mathrm{cp}})
		+\frac{\alpha_{\mathbf k,4}}{6}(\omega - E_{\mathrm{cp}})^2=0$. 
		Since near the cusp $\mathrm{Disc}[\partial_\omega A] \sim -\frac{2\alpha_{\mathbf{k},2}\alpha_{\mathbf{k},4}}{3}>0$, implying $\alpha_{\mathbf{k},2}\alpha_{\mathbf{k},4}<0$, the two solutions straddle $E_{\mathrm{cp}}$, making it correspond to a dip in $A(\mathbf{k},\omega)$.
		
		As a result, the Fermi surface is interrupted upon crossing the cusp, giving rise to a Fermi arc [Fig.~A1(d)]. By plotting the negative second derivative of the spectral function with respect to frequency, the extrema are strongly enhanced: peaks yield positive intensity, forming a clear red arc that remains open. Reference spectra taken at a point on the arc and one outside [Fig.~A1(b),(c)] further confirm this behavior. Notably, the red arc terminates precisely at the blue cusp point, demonstrating that the presence of a cusp necessarily implies the emergence of Fermi arcs.
		
		\section{Appendix B. The effect of fluctuation}
		\subsection{single band case}
		We begin by considering both temporal and spatial fluctuations of the order parameter, treating their combined effect as a spatiotemporally correlated disorder, $V(t)=\int d\mathbf r \Delta(\mathbf r,t)c^\dagger_{\mathbf r}c_{\mathbf r}$, where $\overline{\Delta(\mathbf r,t)}=0,\ \overline{\Delta(\mathbf{r}_i,t_i)\Delta(\mathbf{r}_j,t_j)}=\Delta_0^2 e^{-|\mathbf{r}_i-\mathbf{r}_j|^2/\xi_T^2}e^{-(t_i-t_j)^2/\xi_t^2}$. Treating $V(t)$ as the interaction term, we introduce the time-evolution operator in the interaction picture, $S = T\exp\!\Big[-\,i\!\int_{-\infty}^{\infty}\! dt\, V(t)\Big]$.
		Expanding the exponential in powers of $V(t)$, transforming to momentum–frequency space, and applying Wick’s theorem yield the standard Feynman-diagram expansion. Each vertex corresponds to a single scattering event from the fluctuating order field:
		\begin{align}
			\Delta_{\mathbf k' \mathbf k}^{\omega' \omega}=\tikz[baseline={(current bounding box.center)}]{
				\node[cross out, draw, minimum size=6pt, inner sep=0pt] at (0,0) {};
				\draw[midin]  (0,0) -- (2,0)
				node[above,pos=0.5] {$(\mathbf k,\omega)$};
				\draw[midout] (0,0) -- (-2,0)
				node[above,pos=0.5] {$(\mathbf k',\omega')$};
			}=(-i)\int dt \int d \mathbf r\; \Delta(\mathbf r,t)e^{-i(\mathbf k-\mathbf k‘)\cdot \mathbf r+i(\omega-\omega’)t},
		\end{align}
		Because both spatial and temporal fluctuations are included, energy and momentum are not conserved at individual vertices. Upon performing the ensemble average, the single-vertex contribution vanishes, and the leading nonzero term arises from diagrams containing two vertices. The resulting ensemble-averaged self-energy reads
		\begin{align}
			\Sigma(\mathbf k,\omega;\mathbf k',\omega')=\int  \frac{d\nu}{2\pi} \frac{d^2\mathbf q}{(2\pi)^2} \frac{\overline{\Delta_{k'q}^{\omega'\nu}\Delta_{qk}^{\nu \omega}}}{\nu-\epsilon_{\mathbf{q}}+i\eta}
		\end{align}
		After ensemble averaging, both spatial and temporal translation symmetries are restored, yielding $\Sigma(\mathbf k,\omega;\mathbf k',\omega')=\Sigma(\mathbf k,\omega)\delta(\mathbf k'-\mathbf k)\delta(\omega'-\omega)$, where
		\begin{align}
			\Sigma(\mathbf k,\omega)=\sqrt{2\pi}\xi_t 2\pi \xi_T^2 \Delta_0^2 \left(\int \frac{d\nu}{2\pi} \frac{d^2 \mathbf q}{(2\pi)^2}\frac{e^{-\frac{\xi_T^2}{2}(\mathbf{k-Q-q})^2}e^{-\frac{\xi_t^2}{2}\nu^2}}{\omega+\nu-\epsilon_{\mathbf{q}}+i\eta}
			+\int \frac{d\nu}{2\pi} \frac{d^2 \mathbf q}{(2\pi)^2}\frac{e^{-\frac{\xi_T^2}{2}(\mathbf{k+Q-q})^2}e^{-\frac{\xi_t^2}{2}\nu^2}}{\omega+\nu-\epsilon_{\mathbf{q}}+i\eta}\right)
		\end{align}
		This expression indicates that the dominant contribution arises from electronic states separated by $\mathbf{Q}$.
		Since the Gaussian factor strongly suppresses contributions far from $\mathbf{q} = \mathbf{k} \pm \mathbf{Q}$, we may approximate the integral by expanding around these momenta:
		\begin{align}
			\Sigma(\mathbf k,\omega)=\sqrt{2\pi}\xi_t 4\pi \xi_T^2 \Delta_0^2\int \frac{d\nu}{2\pi} \frac{d^2 \mathbf q}{(2\pi)^2}\frac{e^{-\frac{\xi_T^2}{2}\mathbf{q}^2}e^{-\frac{\xi_t^2}{2}\nu^2}}{\omega+\nu-\epsilon_{\mathbf{k+Q}}-v_{\mathbf{k+Q}}\cdot \mathbf{q}+i\eta}
		\end{align}
		where $\mathbf{v}_{\mathbf{k}} = \nabla_{\mathbf{k}}\epsilon_{\mathbf{k}}$ is the Fermi velocity.
		The imaginary part of the self-energy is then
		\begin{align}
			\mathrm{Im}\Sigma(\mathbf k,\omega)&=-\sqrt{2\pi}\xi_t 4\pi \xi_T^2 \Delta_0^2\int_{-\infty}^{\infty}\frac{d\nu}{2\pi}e^{-\frac{\xi_t^2}{2}\nu^2} \int_0^\infty \frac{q dq}{(2\pi)^2}\int_0^{2\pi}d\theta e^{-\frac{\xi_T^2}{2}q^2}\delta(\omega+\nu-\epsilon_{\mathbf{k+Q}}-q v_{\mathbf{k+Q}} \cos\theta)\\
			&=\sqrt{2\pi}\xi_t\frac{\xi_T \Delta_0 ^2 }{v_\mathbf{k+Q}}\int_{-\infty}^{\infty}\frac{d\nu}{2\pi}e^{-\frac{\xi^2(\omega+\nu -\epsilon_{\mathbf{k+Q}})^2}{2  v_\mathbf{k+Q}^2}}e^{-\frac{\xi_t^2}{2}\nu^2}\\
			&=\Delta_0 ^2\frac{\xi_T}{v_\mathbf{k+Q}}\sqrt{\frac{\xi_t^2}{\xi_t^2+\frac{\xi_T^2}{v_\mathbf{k+Q}^2}}}e^{-\frac{\xi_T^2\xi_t^2/v_\mathbf{k+Q}^2}{2(\xi_T^2/v_\mathbf{k+Q}^2+\xi_t^2)}(\omega-\epsilon_{\mathbf{k+Q}})^2}\\
			&=\Delta_0 ^2\frac{\xi}{v_\mathbf{k+Q}}e^{-\frac{\xi^2}{2v_\mathbf{k+Q}^2}(\omega-\epsilon_{\mathbf{k+Q}})^2}
		\end{align}
		with $\xi=\xi_T\sqrt{\frac{\xi_t^2}{\xi_t^2+\frac{\xi_T^2}{v_\mathbf{k+Q}^2}}}$. 
		This result shows that $\mathrm{Im}\Sigma(\mathbf{k},\omega)$ follows a Gaussian form.
		Replacing the Gaussian by a Lorentzian profile and applying the Kramers–Kronig relation yields the final expression for the self-energy:
		\begin{align}
			\Sigma(\mathbf{k}, \omega) = \frac{\Delta_0^2}{\omega - \epsilon_{\mathbf{k} + \mathbf{Q}} + i v_{\mathbf{k} + \mathbf{Q}}/\xi}
		\end{align}
		When $v_{\mathbf{k}+\mathbf{Q}} \xi_t \gg \xi_T$, we have $\xi \approx \xi_T$, which makes this expression equivalent to Eq.~(\ref{eq:SelfEnergy}) in the main text.
		This indicates that the effect of temporal fluctuations is effectively absorbed into a renormalization of the spatial correlation length.
		Therefore, it is sufficient to consider only spatial fluctuations in the main analysis while neglecting temporal ones.
		\subsection{two bands case}
		In certain cases, the spin-density-wave (SDW) order opens a gap between different bands rather than within a single band. For example, in iron-based systems with model Hamiltonian [the same model in Eq.~\ref{Eq:LatticeModel}], 
		\begin{align}
			H_0=
			\begin{aligned}
				(d_{x\mathbf{k}}^\dagger, d_{y\mathbf{k}}^\dagger)
				h_0(\mathbf k)
				\left(
				\begin{aligned}
					d_{x\mathbf{k}}\\
					d_{y\mathbf{k}}
				\end{aligned}
				\right)
			\end{aligned}
			, \mathrm{with}\  
			\begin{aligned}
				h_0(\mathbf k)=	\left(
				\begin{matrix}
					\varepsilon_{x\mathbf{k}} -\mu& \varepsilon_{x y\mathbf{k}}\\
					\varepsilon_{x y\mathbf{k}}&\varepsilon_{y\mathbf{k}}-\mu
				\end{matrix}
				\right),
			\end{aligned}
		\end{align}
		the $d_x$ and $d_y$ orbitals hybridize into two bands $\epsilon_{\pm,\mathbf{k}}$. The SDW order then couples $\epsilon_{-,\mathbf{k}}$ and $\epsilon_{+,\mathbf{k+Q}}$.
		To account for fluctuation effects in this situation, we assume for simplicity that the order parameter carries equal weight on different orbitals. The fluctuating order can then be expressed as
		$V = \int d\mathbf{r} \left( e^{i\mathbf{Q}\cdot\mathbf{r}} \Delta(\mathbf{r},t)\, (d^\dagger_{x\mathbf{r}} d_{x\mathbf{r}} + d^\dagger_{y\mathbf{r}} d_{y\mathbf{r}})+h.c. \right)$.
		Consequently, each vertex in the Feynman diagram of Eq.~(A1) becomes a $2\times2$ matrix rather than a scalar.
		\begin{align}
			\Delta_{\mathbf k' \mathbf k}^{\omega' \omega}=(-i)\int dt \int d \mathbf r\; \Delta(\mathbf r,t)e^{-i(\mathbf k-\mathbf k‘)\cdot \mathbf r+i(\omega-\omega’)t}U^\dagger_\mathbf{k'}U_\mathbf{k},
		\end{align}
		Here $U_{\mathbf{k}}$ is the unitary matrix that diagonalizes $h_0(\mathbf{k})$, satisfying
		$U_{\mathbf{k}}^\dagger h_0(\mathbf{k}) U_{\mathbf{k}} = \mathrm{diag}(\epsilon_{-,\mathbf{k}}, \epsilon_{+,\mathbf{k}})$.
		The ensemble-averaged self-energy then takes the form
		\begin{align}
			\Sigma(\mathbf k,\omega)=\int \frac{d\nu}{2\pi} \frac{d^2 \mathbf q}{(2\pi)^2}e^{-\frac{\xi_T^2}{2}\mathbf{q}^2}e^{-\frac{\xi_t^2}{2}\nu^2}
			U^\dagger_{\mathbf{k}}U_{\mathbf{k+Q+q}}
			\begin{aligned}
				\left(
				\begin{matrix}
					\frac{\sqrt{2\pi}\xi_t 4\pi \xi_T^2 \Delta_0^2}{\omega+\nu-\epsilon_{-,\mathbf{k+Q}}-v_{-,\mathbf{k+Q}}\cdot \mathbf{q}+i\eta}& 0\\
					0&\frac{\sqrt{2\pi}\xi_t 4\pi \xi_T^2 \Delta_0^2}{\omega+\nu-\epsilon_{+,\mathbf{k+Q}}-v_{+,\mathbf{k+Q}}\cdot \mathbf{q}+i\eta}
				\end{matrix}
				\right)
			\end{aligned}
			U^\dagger_{\mathbf{k+Q} + \mathbf{q}}U_\mathbf{k}
		\end{align}
		Owing to the Gaussian factor $e^{-\xi_T^2 \mathbf{q}^2 / 2}$, the dominant contribution to the integral arises from momenta near $\mathbf{q} = 0$, so that $U^\dagger_{\mathbf{k+Q}+\mathbf{q}} \approx U^\dagger_{\mathbf{k+Q}}$.
		Under this approximation, each matrix element can be integrated independently, yielding the final expression for the self-energy,
		\begin{align}
			\Sigma(\mathbf k,\omega)=
			\left(
			\begin{matrix}
				\frac{\Delta_0^2 \cos^2(\frac{\theta_\mathbf{k}-\theta_{\mathbf{k+Q}}}{2})}{\omega - \epsilon_{-,\mathbf{k} + \mathbf{Q}} + i v_{-,\mathbf{k} + \mathbf{Q}}/\xi}+\frac{\Delta_0^2 \sin^2(\frac{\theta_\mathbf{k}-\theta_{\mathbf{k+Q}}}{2})}{\omega - \epsilon_{+,\mathbf{k} + \mathbf{Q}} + i v_{+,\mathbf{k} + \mathbf{Q}}/\xi}& \frac{1}{2}(\frac{\sin(\theta_{\mathbf{k}}-\theta_{\mathbf{k+Q}})}{\omega - \epsilon_{+,\mathbf{k} + \mathbf{Q}} + i v_{+,\mathbf{k} + \mathbf{Q}}/\xi}-\frac{\sin(\theta_{\mathbf{k}}-\theta_{\mathbf{k+Q}})}{\omega - \epsilon_{-,\mathbf{k} + \mathbf{Q}} + i v_{-,\mathbf{k} + \mathbf{Q}}/\xi})\\
				\frac{1}{2}(\frac{\sin(\theta_{\mathbf{k}}-\theta_{\mathbf{k+Q}})}{\omega - \epsilon_{+,\mathbf{k} + \mathbf{Q}} + i v_{+,\mathbf{k} + \mathbf{Q}}/\xi}-\frac{\sin(\theta_{\mathbf{k}}-\theta_{\mathbf{k+Q}})}{\omega - \epsilon_{-,\mathbf{k} + \mathbf{Q}} + i v_{-,\mathbf{k} + \mathbf{Q}}/\xi})&\frac{\Delta_0^2 \cos^2(\frac{\theta_\mathbf{k}-\theta_{\mathbf{k+Q}}}{2})}{\omega - \epsilon_{+,\mathbf{k} + \mathbf{Q}} + i v_{+,\mathbf{k} + \mathbf{Q}}/\xi}+\frac{\Delta_0^2 \sin^2(\frac{\theta_\mathbf{k}-\theta_{\mathbf{k+Q}}}{2})}{\omega - \epsilon_{-,\mathbf{k} + \mathbf{Q}} + i v_{-,\mathbf{k} + \mathbf{Q}}/\xi}
			\end{matrix}
			\right)
		\end{align}
		with $\sin\theta_{\mathbf{k}}=\frac{\varepsilon_{xy\mathbf{k}}}{\sqrt{\varepsilon_{xy\mathbf{k}}^2+(\varepsilon_{x \mathbf{k}}-\varepsilon_{y \mathbf{k}})^2/4}}$. 
		
		As shown in Fig.~\ref{figsm:2}(a), when focusing on the physics near the $\Gamma$ point, the relevant couplings involve $\epsilon_{-,\mathbf{k}}$ and $\epsilon_{\pm,\mathbf{k+Q}}$. The corresponding Green’s function for the three-band model reads $G_{--}^{\mathrm{3band}}(\mathbf k,\omega)=\left(\omega-\epsilon_{-,\mathbf k}-\frac{\Delta_0^2 \cos^2(\frac{\theta_\mathbf{k}-\theta_{\mathbf{k+Q}}}{2})}{\omega - \epsilon_{-,\mathbf{k} + \mathbf{Q}} + i v_{-,\mathbf{k} + \mathbf{Q}}/\xi}-\frac{\Delta_0^2 \sin^2(\frac{\theta_\mathbf{k}-\theta_{\mathbf{k+Q}}}{2})}{\omega - \epsilon_{+,\mathbf{k} + \mathbf{Q}} + i v_{+,\mathbf{k} + \mathbf{Q}}/\xi}\right)^{-1}$. 
		However, near $\Gamma$, only $\epsilon_{-,\mathbf{k}}$ and $\epsilon_{+,\mathbf{k+Q}}$ lie close to the Fermi level, while $\epsilon_{-,\mathbf{k+Q}}$ is far away, $|\epsilon_{-,\mathbf{k+Q}}-\mu|\gg1$, and thus has negligible influence.
		Neglecting this high-energy band yields an effective two-band Green’s function, $G_{--}^{\mathrm{2band}}(\mathbf k,\omega)=\left(\omega-\epsilon_{-,\mathbf k}-\frac{\Delta_0^2 \sin^2(\frac{\theta_\mathbf{k}-\theta_{\mathbf{k+Q}}}{2})}{\omega - \epsilon_{+,\mathbf{k} + \mathbf{Q}} + i v_{+,\mathbf{k} + \mathbf{Q}}/\xi}\right)^{-1}$ which is equivalent to replacing $\Delta_0$ in the main-text formula by a momentum-dependent coupling $\Delta_\mathbf{k}=\Delta_0\sin\frac{\theta_\mathbf{k}-\theta_{\mathbf{k+Q}}}{2}$.
		
		\begin{figure}[h]
			\centering
			\includegraphics[width=\linewidth]{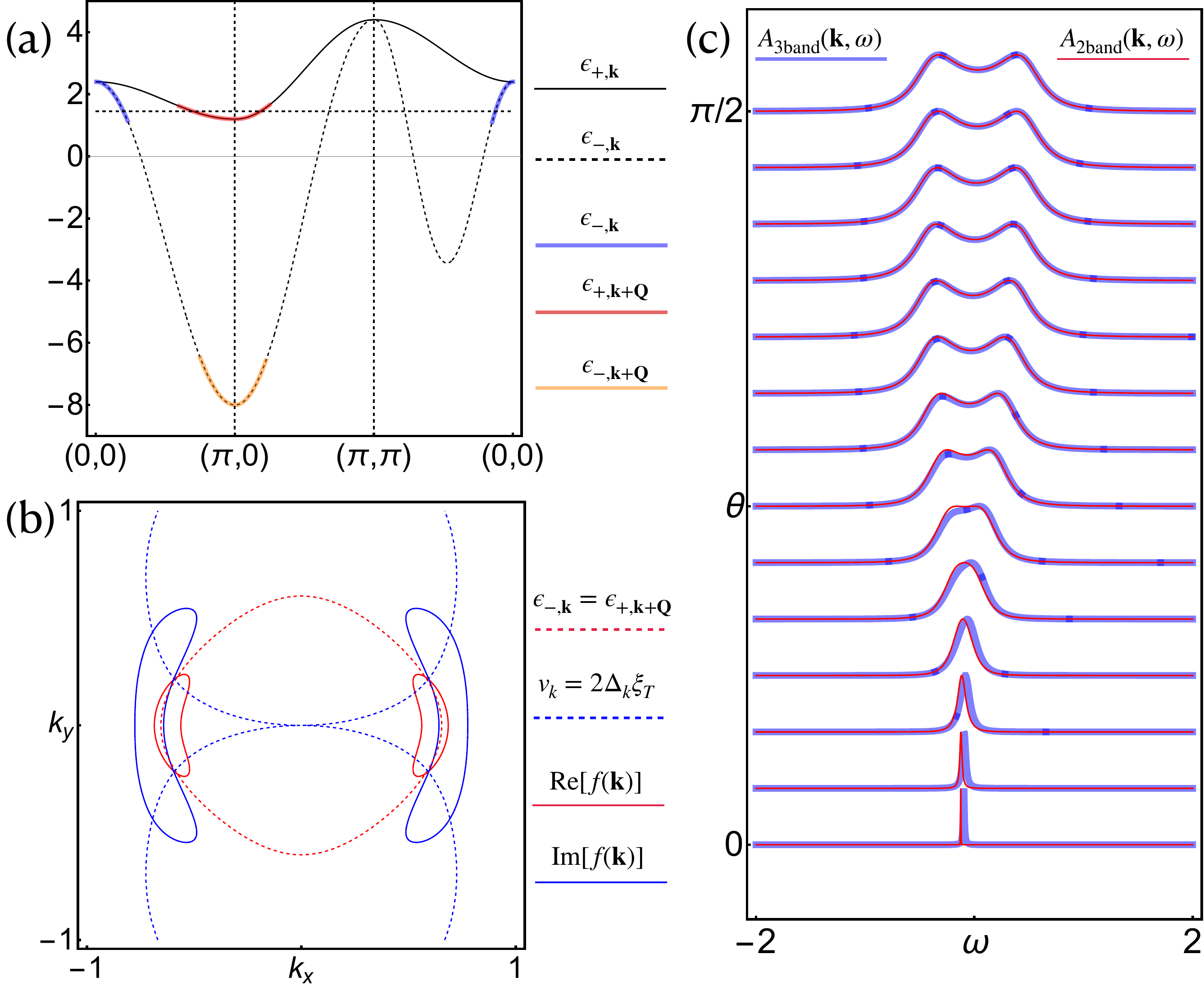}
			\caption{
				Illustration based on the model in Eq.~\ref{Eq:LatticeModel} with parameters $(t_1, t_2, t_3, t_4, \mu, \Delta_0, \xi)=(-0.9, 1.4, -0.85, -0.85, 1.45, 0.5, 1.845)$.
				(a) Black solid and dashed lines denote the hybridized bands $\epsilon_{\pm,\mathbf{k}}$; colored segments indicate the portions coupled by the order parameter.
				(b) Red and blue solid lines show the EP conditions for the three-band model, with their intersection marking the EP. The red and blue dashed lines correspond to the two-band approximation, where their intersection gives the EP position.
				(c) Comparison of the Fermi-surface geometry near the nesting line $\epsilon_{-,\mathbf{k}}=\epsilon_{+,\mathbf{k+Q}}$, obtained by traversing $\mathbf{k}$ counterclockwise from the positive $k_x$ axis for both the three-band and two-band models.
			}
			\label{figsm:2}
		\end{figure}
		
		To verify the validity of this approximation, we compare the EP conditions and spectral functions obtained from $G_{--}^{(3\mathrm{band})}(\mathbf{k},\omega)$ and $G_{--}^{(2\mathrm{band})}(\mathbf{k},\omega)$.
		Since $G_{--}^{(2\mathrm{band})}$ corresponds to the Green’s function used in the main text, its EPs are determined by the same condition as in Eq.~(\ref{Eq:DeltaEqgamma}) and Eq.~(\ref{Eq:DpLine}) by replacing $\Delta_0$ with $\Delta_{\mathbf k}$, corresponding to the intersection of the red and blue dashed lines in Fig.~A2(b).
		For the three-band Green’s function, EPs are defined by the simultaneous solutions of $(G_{--}^{(3\mathrm{band})})^{-1}=0$ and $\partial_\omega(G_{--}^{(3\mathrm{band})})^{-1}=0$. These two equations yield a pair of complex polynomial constraints whose resultant $f(\mathbf{k})$ must vanish. Thus, the EP condition reduces to finding common zeros of $\mathrm{Re}[f(\mathbf{k})]=0$ and $\mathrm{Im}[f(\mathbf{k})]=0$, corresponding to the intersection of the red and blue solid lines in Fig.~\ref{figsm:2}(b).
		The two sets of EPs nearly coincide, confirming that the two-band Green’s function faithfully captures the essential EP physics.
		Furthermore, along the nesting line $\epsilon_{-,\mathbf{k}}=\epsilon_{+,\mathbf{k+Q}}$, we compare the spectral functions of the two models by scanning $\mathbf{k}$ clockwise from the $k_x$ axis. As shown in Fig.~\ref{figsm:2}(c), both spectra exhibit almost identical line shapes near the Fermi surface, confirming that the spectral cusp and EP structure obtained from $G_{--}^{(2\mathrm{band})}$ accurately describe the full three-band system.
		
		\section{Appendix C. Derivation of Eq.~(\ref{Eq:decay}) in the main text}
		\subsection{signature of Tr-ARPES through ultrafast probe}
		Following Refs.~\cite{JKF_PRL_TrARPES,JKF_PRX_2018,nonequilibrium_textbook_stefanucci2013}, we consider a two-dimensional system described by the Hamiltonian $H_{\mathrm{system}}$, subject to a time-dependent pump field $H_{\mathrm{pump}}(t)$.
		A probe pulse is then applied, described by
		\begin{align}
			\label{eqHprobe}
			\hat{H}_{probe}(t)=s(t) \sum_{\mathbf{k} \sigma, \nu}\left(M_{\mathbf{k} \sigma \nu}e^{-i\omega_qt} \hat{f}_{\mathbf{k} \sigma}^{\dagger} \hat{c}_{\mathbf{k}_{\|} \nu}+M_{\mathbf{k} \sigma \nu}^*e^{i\omega_qt} \hat{c}_{\mathbf{k}_{\|} \nu}^{\dagger} \hat{f}_{\mathbf{k} \sigma}\right),
		\end{align}
		where $\hat{f}_{\mathbf{k}\sigma}^\dagger$ creates a photoelectron with three-dimensional momentum $\mathbf{k}$, and $\hat{c}_{\mathbf{k}_{\parallel}\nu}^\dagger$ creates a Bloch electron in the $\nu$-th band with in-plane momentum $\mathbf{k}_{\parallel}$.
		The coupling matrix element
		$M_{\mathbf{k} \sigma \nu} \equiv \frac{1}{c}\left\langle f_{\mathbf{k} \sigma}\right|(\hat{\mathbf{p}} \cdot \boldsymbol{A})\left|\mathbf{k}_{\|} \nu\right\rangle$ describes the dipole interaction between the photoelectron and the Bloch state.
		The vector potential $\mathbf{A}(t)=s(t)\mathbf{A}$ represents the light field, where $s(t)$ is the envelope function of the pulse amplitude. 
		The resulting photoemission current is given by
		\begin{align}
			\langle \mathbf J_d\rangle(t)-\langle \mathbf J_d\rangle(0)&=\frac{\hbar \mathbf k_e}{m_e}P(t)
			\\ 
			P(t)&=\sum_{\nu_1,\nu_2}\left(\sum_{\sigma_e}M_{k_e,\sigma_e,\nu_1}^*M_{k_e,\sigma_e,\nu_2} \right)\int_{t_0}^t dt_2\int_{t_0}^t dt_1 s(t_1)s(t_2)e^{i\omega(t_2-t_1)}G^<_{k_{e,||},\nu_1,\nu_2}(t_1,t_2)
		\end{align}
		where $G^<_{k_{e,||},\nu_1,\nu_2}(t_1,t_2)$ is the non-equilibrium lesser Green’s function.
		For an ultrafast probe centered at time $t=\tau$, one may approximate the envelope as $s(t)=\delta(t-\tau)$.
		Assuming that only a single band is excited, the photocurrent simplifies to
		\begin{align}
			P(\infty)=G_{\mathbf k,\nu,\nu}^<(\tau,\tau)=\langle c_\mathbf{k}^\dagger(\tau) c_\mathbf{k}(\tau)\rangle
		\end{align}
		showing that the photoemission intensity is proportional to the instantaneous occupation of momentum $\mathbf{k}$ at delay time $\tau$.
		
		\subsection{the occupation number after a ultrafast pump}
		Before probing the system at delay time $\tau$, we first excite it with an ultrafast pump pulse at $t=0$, modeled as $H_{pump}(t)=\delta(t)O$. For a laser pump, the perturbation operator takes the form $O=\sum_\mathbf{k} Y_\mathbf{k} c_{\mathbf{k+q_p}}^\dagger c_{\mathbf{k}}$, where $\mathbf{q}_p$ denotes the photon momentum.
		Assuming the initial ground state $|\psi_0\rangle$, the time-dependent momentum occupation is given by
		\begin{align}
			\langle c_\mathbf{k}^\dagger(\tau) c_\mathbf{k}(\tau)\rangle=\langle \psi_0| e^{i O}e^{i H\tau}  c_\mathbf{k}^\dagger c_\mathbf{k} e^{-iH\tau}e^{-iO} |\psi_0 \rangle
		\end{align}
		For a weak pump, $|O|\ll 1$, we expand $e^{iO} \approx 1 + iO$ and keep only the leading-order contribution.
		The time-dependent change in the occupation is then
		\begin{align}
			\delta n_\mathbf{k}(t)=& \langle \psi_0| O e^{i H\tau}  c_\mathbf{k}^\dagger c_\mathbf{k} e^{-iH\tau}O |\psi_0 \rangle \label{eq:A1}
		\end{align}
		Because the post-pump dynamics are described by the mean-field Hamiltonian $H$ in Eq.\ref{Eq:H}, Wick’s theorem can be applied to evaluate Eq.\ref{eq:A1}, leading to
		\begin{align}
			\delta n_\mathbf{k}(t)=& |Y_{\mathbf{k}-\mathbf q_p} b|^2  n_0(\mathbf{k} - \mathbf{q}_p) |G^>(\mathbf{k}, t)|^2 - |Y_{\mathbf{k}} b|^2(1 - n_0(\mathbf{k} + \mathbf{q}_p)) |G^<(\mathbf{k}, t)|^2,\label{eq:A2}
		\end{align}
		where $G^<(\mathbf{k}, t)=\langle \psi_0| e^{iHt}c_\mathbf{k}^\dagger e^{-iHt} c_\mathbf{k} |\psi_0\rangle$, $G^>(\mathbf{k}, t)=\langle \psi_0| c_\mathbf{k} e^{iHt}c_\mathbf{k}^\dagger e^{-iHt}  |\psi_0\rangle$ are the equilibrium lesser and greater Green’s functions.
		Eq.~\ref{eq:A2} corresponds to Eq.~(\ref{Eq:decay}) in the main text.
		
	\end{widetext}
	
\end{document}